\documentclass[cits]{PoS}

\newcommand{\be}{\begin{eqnarray}}
\newcommand{\ee}{\end{eqnarray}}

\title{Dipole Straylight Contamination and Low Multipoles}

\ShortTitle{Dipole Straylight Contamination and Low Multipoles}

\author{Alessandro Gruppuso, Carlo Burigana\\
        INAF-IASF Bologna, via Gobetti 101, I-40129 Bologna, Italy\\
        E-mail: \email{gruppuso@iasfbo.inaf.it},
		\email{burigana@iasfbo.inaf.it}}

\author{Fabio Finelli\\
        INAF-IASF Bologna, via Gobetti 101, I-40129 Bologna, Italy \\
        and INAF-OAB, via Ranzani 1, I-40127 Bologna, Italy\\
        E-mail: \email{finelli@iasfbo.inaf.it}}

\abstract{Kinematic dipole signal originates from the motion of the Sun 
with respect to the CMB rest frame.
It is a remarkable signal, two orders of magnitude larger than the observed pattern
of CMB anisotropies.
Therefore the dipole subtraction from the main beam is of primary importance.
But what happens to the dipole signal entering the main spillover 
(i.e. the most relevant far antenna pattern feature, responsible for the 
bulk of the straylight effect)?
We present here an analytical and statistical study of this systematic effect for a
spinning space-mission like {\sc Planck}
\footnote{This work has been done in the framework of the {\sc Planck} LFI activities.}.}


\FullConference{CMB and Physics of the Early Universe\\
		 20-22 April 2006\\
		 Ischia, Italy}

\begin{document}

\section{Analytical Description}
The convolution $I$ of the dipole signal with the
main spillover 
\footnote{It is the main feature of the antenna pattern
far from the main beam.} can be written as:
\be I = \int d\Omega \, T_{1 m} Y_1^m (\theta,
\varphi) B_{SL}(\theta, \varphi) \, , \label{conv}
\ee
where
$d\Omega = d \theta \sin \theta \, d \varphi $ 
is the solid angle element, $\theta \in \left[0,\pi \right]$ 
is the colatitude, $\varphi \in \left[0, 2 \pi\right.\left[\right.$
is the longitude, and where the sum on $m$
over $-1,0,1$ is understood. $T_{1 m}$ are the coefficients of the
expansion of the dipole 
\footnote{We use the symbol $T_{\ell m}$
in order to make clear that the dimensionality is a temperature ($^{\circ} \! K$).
Moreover we work in a frame centered on the spacecraft with axes pointing 
fixed (far away) stars.
In this frame the dipole vector is constant while the direction of the main spillover varies (it rotates of $2 \pi$ per year in a nominal scanning strategy).} 
on the spherical harmonics basis $Y_1^m
(\theta, \varphi)$ and $B_{SL}(\theta, \varphi)$ is the beam response
representing the shape of the main spillover in the $(\theta,
\varphi)$-plane. In this notation $B_{SL}$ is normalized to the
whole beam integrated response, dominated by the contribution
of the main beam 
$
\int_{4\pi} d\Omega B \simeq \int_{{\rm main~beam}} d\Omega B \simeq 2\pi \sigma_b^2
$
where $\sigma_b={\rm FWHM}/\sqrt{8{\rm ln}2}$.
Notice that $I$ is a function that depends on the 
istantaneous orientation of the main spillover.

Of course eq.~(\ref{conv}) is general and exact. 
Any specific approximation of $B_{SL}$ (inspired by experiments) 
will introduce a certain degree of uncertainty with respect to 
the real case. 
Our current approximation for $B_{SL}$ is the {\it Top Hat} \cite{report2004,report2005}:
\be B_{SL} (\theta, \varphi) = \, f_{SL}
\, \Delta(\, \theta ,\, \theta_{ms} - \Delta _{\theta <},\, \theta_{ms}
+ \Delta_{\theta >}) \Delta(\, \varphi,\, \varphi_{ms} -
\Delta_{\varphi <}, \, \varphi_{ms} + \Delta_{\varphi >}) \, ,
\label{windowSL}
\ee
where $f_{SL}=p/\Omega_{ms}$ is a constant ($p$ is the ratio between the power entering the main spillover and the power entering the main beam, $\Omega_{ms}$ is the solid angle subtented by the main spillover) and $\Delta (a,b,c) \,=\, S(a - b) \, - \, S(a - c)$ with $S(x)$ representing the step function (or Heavyside function) that takes the value $1$ for $x \ge 0$ and the value $0$ otherwise. Eq.~(\ref{windowSL}) is
just an asymmetric rectangular box, in the $(\theta ,\varphi)$-plane,
centered around the point $(\theta_{ms},\varphi_{ms})$
and with sides of length $\Delta_{\theta >} + \Delta_{\theta <}$ and
$\Delta _{\varphi >} + \Delta _{\varphi <}$. 
Notice that the point
$(\theta_{ms},\varphi_{ms})$ is the pointing direction of
the main spillover. 
With $\epsilon$ and $\delta$ implicitely defined by $\Delta_{\theta >} = \Delta_{\theta } + \delta $ and $\Delta_{\varphi >} = \Delta_{\varphi } + \epsilon $ where $\Delta_{\theta <} = \Delta_{\theta }$ and $\Delta_{\varphi <} = \Delta_{\varphi }$,
we obtain the final expression for the convolution (\ref{conv}):
\begin{eqnarray}
& & I^{TH} / f_{SL} = 
{T_{10}\over 2} \sqrt{3 \over 4 \pi} \sin \left( \delta + 2 \Delta_{\theta}\right)
\sin \left( \delta + 2 \,\theta_{ms}\right)
(2 \Delta_{\varphi } + \epsilon ) -\nonumber \\
& & 4 \sqrt{3 \over 8 \pi} \left( \Delta_{\theta} + {\delta \over 2} -
\cos \left( \delta + 2 \, \theta_{ms} \right)
\, {\sin \left( \delta + 2 \, \Delta_{\theta}\right)\over 2} \right)
Re \left[ T_{11} e^{i (\varphi_{ms}+\epsilon/2)}\right]
\sin \left( \Delta_{\varphi} + {\epsilon \over 2}\right)
\, , \label{convolution}
\end{eqnarray}
where the label $^{TH}$ stands for {\it Top Hat}.
%

During the rotation of the main beam (assumed, for simplicity, at
$90 ^\circ$ from the spin axis) in each scan circle, the main spillover, if not lying 
on the spin axis, draws a cone.
We define $\alpha$ the angle between the main spillover direction and the spin axis
\footnote{For simplicity we assume here the main spillover centre
located on the plane defined by the spin axis
and the beam centre direction.}.
%
%
%
%
%
%
%
The total signal received by the satellite, is the sum 
of the two following terms:
$
T(\theta , \varphi) = T_{MB}(\theta ,\varphi) + I_{SL}(\theta,\varphi)
$
where $T_{MB}$ is the signal entering the main beam where the
dipole has been subtracted away, whereas $I_{SL}$ is the signal
due to the dipole entering the main spillover.
In the case of a {\sc Planck} nominal scanning strategy 
\cite{ss_document}, for the first (upper signs) and the second (lower signs) survey, $I_{SL}$ 
is given by 
\be I^{(I/II)}_{SL}(\theta,\varphi)= \left\{
\begin{array}{c} I^{TH}( \pi /2 - \alpha \cos \theta , \varphi \pm \pi/2 \mp \alpha \sin \theta )  \;
\mbox{for} \; \; \; 0 < \varphi < \pi  \, , \\  \\
I^{TH}(\pi /2 - \alpha \cos \theta , \varphi \mp \pi/2 \pm \alpha \sin \theta ) 
 \; \mbox{for} \; \; \; \pi < \varphi < 2 \pi
\, .
\end{array}\right. \!\!\!\!\!\!\!\! \label{ISL1and2survey}
\ee
The shift in the definition of the arguments in $I_{SL}$ (during either 
the first or the second survey) comes from the fact that
when the main beam rotates from North to South the main spillover
is shifted of $-\pi/2$ plus a small correction proportional to $ \alpha$
(due to the non perfect alignment of the main spillover with the spin 
axis), while when the main beam rotates
from South to North the main spillover is shifted of $ +\pi/2$ minus
a small correction again proportional to $\alpha$.
Notice that $(\theta , \varphi)$ is the direction of pointing of the main beam.

\section{General Analytical Results}
\label{generalresults}
We set now $T_{MB}=0$ and take into account only the straylight signal,
\be
T^{SL}_{\ell m} = \int d\Omega \, I_{SL}(\theta, \varphi) \,
Y_{\ell m}^{\star}(\theta , \varphi)
\, , \label{TSL}
\ee
because we want to analyze only the map due to the Dipole Straylight Contamination (DSC).
Some general properties can be found \cite{dipolepaper}, once performed an expansion for small $\alpha$: $
T_{\ell m}^{SL} = T_{\ell m}^{(0)} + \alpha T_{\ell m}^{(1)} + {\cal O}(\alpha^2)
$.
In particular for an odd number of surveys to the leading order in $\alpha$ only the even multipoles survive [i.e. $T_{\ell m}^{(0)} \neq 0$ for even $\ell$ (and $m$)] while to the linear order only the $\ell =1$ term turns on.
For an even number of surveys DSC vanishes except for $\ell=1$,
$\hat T _{\ell m}^{SL} = (T _{\ell m}^{(1st)} + T _{\ell m}^{(2nd)})/2 =
0 + \alpha T_{\ell m}^{(1)} + {\cal O}(\alpha^2)
$.
These qualitative features have been tested and extended by numerical simulations 
\cite{dipolepaper}.
As an example, we show the analytical expressions derived for the dipole 
\begin{eqnarray}
T^{SL}_{1 0} = 2 \sqrt{4 \pi / 3}  c_1 \alpha
\, , \\ 
T_{1 \pm 1}^{SL} = {(1 / 2)}\sqrt{8 \pi/ 3} \left( \pm d_1 + i d_2 \right)
(c_2 + c_3) \alpha
\, ,
\end{eqnarray}
and the non vanishing coefficients for the 
quadrupole
\be T^{SL}_{2 \pm 2} = - \left({4 / 3}\right)^2 \sqrt{{15 / 32 \pi}}
\left( d_1 \pm 2 i d_2 \right) \left( c_2 + c_3 \right)
\, ,
\ee
where 
$ c_1 = \sqrt{3 / 4 \pi} \, f_{SL} \Delta \, \sin (2 \Delta) \, T_{10} $ , 
$c_2 = 4 \sqrt{3 / 8 \pi} \, f_{SL}\, \Delta$ ,
$c_3 = 4 \sqrt{3 / 8 \pi}\, f_{SL} \, \sin (2 \Delta) /2$ , 
$d_1 = \sin \Delta \, Re \left[ T_{11}\right] $ ,
$d_2 = \sin \Delta \, Im \left[ T_{11}\right] $ ,
with $\Delta_{\theta}= \Delta_{\varphi}=\Delta$ and $\epsilon = \delta =0$.

\section{Dipole and Quadrupole Analysis}
\label{dipolequadrupole}
Section \ref{generalresults} shows that when the main spillover direction is different from the spin axis (i.e. $\alpha \neq 0$) then the {\em Dipole} itself is modified. 
This has an impact on the calibration with amplitude related to the value of $p$.
For $\Delta_{\theta}= \Delta_{\varphi}=\Delta= \pi /10$, $\alpha = \pi/18$ and $p=1/100$,
we estimate that $
T_{10}^{SL} = -2.1 \mu K \, ,
T_{11}^{SL} = (7.7 \pm 1.0 i) \mu K $.
This leads to 
$ C_1^{SL} = 44.5 \mu K^2 $.

\begin{figure}
\begin{tabular}{cc}
\centering
\includegraphics[width=6.7cm]{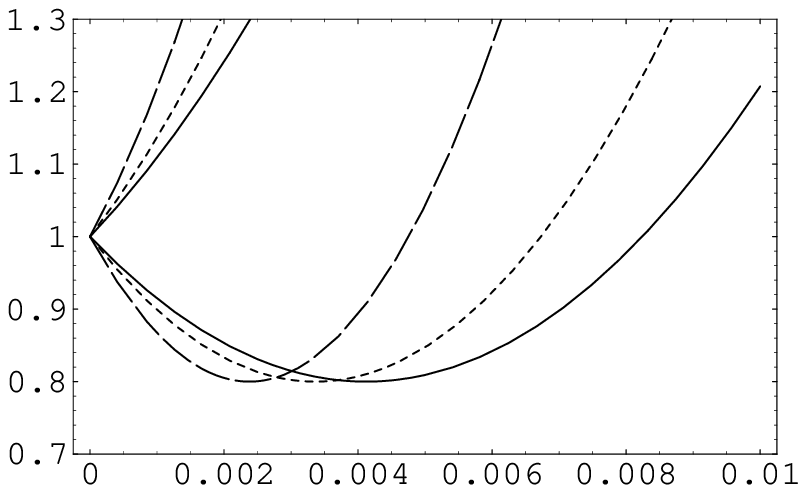}
&
\includegraphics[width=6.7cm]{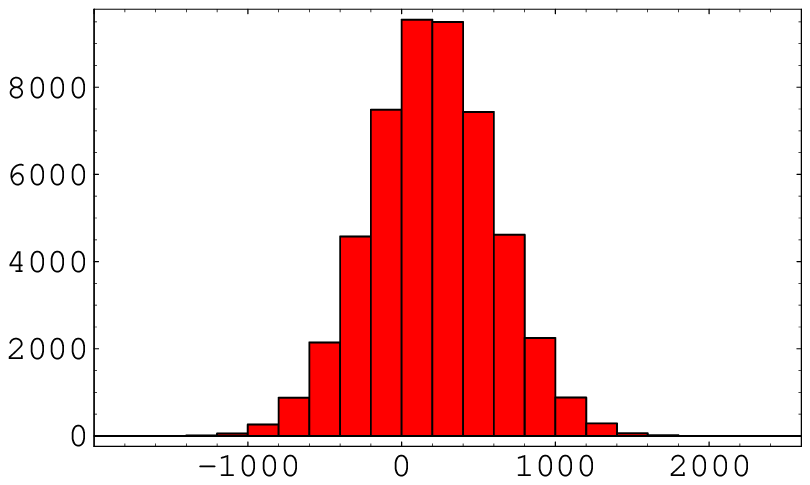}
\end{tabular}
\caption{Left Panel: $y=y(F_{SL})$ with $B>0$ (lower curves).
Solid line is for $C_2^{SKY} = 1500 \mu K^2$, dotted line is for
$C_2^{SKY} = 1000 \mu K^2$ and dashed line is for $C_2^{SKY} = 500 \mu K^2$.
We report also the case of $B<0$ (upper curves)
which does not imply a quadrupole decreasing but only a quadrupole increasing.
Right Panel: Distribution of the DSC, i.e. $C_2-C_2^{SKY}$.
See also the text.}
\label{figura}
\end{figure}

The next multipole that is modified by DSC is the {\em Quadrupole}.
We can write the observed quadrupole $C_2$ as
$
C_2 (F_{SL})= C_2^{SKY} - {4 \over 5} B \, F_{SL} + {2 \over 5} A \, F_{SL}^2
\, ,
$
where 
$C_2^{SKY}$ is the intrinsic quadrupole (that would be observed if the DSC would be absent),
$
B \equiv Re \left[ T_{11}\right] Re
\left[ T_{22}^{SKY} \right] + 2 Im \left[ T_{11} \right] Im \left[
T_{22}^{SKY} \right]$,
$A \equiv \left[{Re\left[
T_{11}\right]}\right]^2 + 4 \, \left[{Im \left[ T_{11}\right]}\right]^2 \, $
and
$F_{SL} =
{ (\sqrt{5}\,p / 3 \pi )} \left[ 1
+ \cos \Delta \, \sin \Delta / \Delta\right]$.
Depending on the sign of $B$ (or in other words on the sign of the real and the imaginary part of $T_{22}^{SKY}$) we have two different kinds of behaviour for the observed quadrupole in terms of $F_{SL}$. It can only increase for $B<0$ but it can also decrease for $B>0$. 
In the left panel of Fig.~\ref{figura} the two branches are plotted for the function $y(F_{SL})=C_2(F_{SL})/C_2^{SKY}$
where, as an example,
it has been set
$ Re \left[ T_{22}^{SKY} \right] \sim \pm (C_2^{SKY}/2)^{1/2}$ and
where it has been considered 
$Im \left[ T_{11} \right] \ll Re \left[ T_{11} \right]$.
In this regime $Im \left[ T_{22}^{SKY} \right]$ does not enter in the estimate.

Since the behaviour of the observed quadrupole depends on the sky realization ($T_{22}^{SKY}$), we present a statistical analysis 
aimed at the computation of the probability of increasing or decreasing of $C_2$.
We have made $50000$ Gaussian distributed extractions of $T_{2 m}^{SKY}$ such that on average $C_{2}^{SKY} \sim \, 1000 \, \mu K^2$ and we have computed the distribution for the observed quadrupole $C_2$ and
for the intrinsic quadrupole. 
In the right panel of Fig.~\ref{figura} we show the plot of the distribution for the pure DSC (i.e. the difference $C_2-C_{2}^{SKY}$).
We note that the distribution of DSC has a Gaussian profile and we compute its mean ($202.3 ~\mu K^2$) and standard deviation ($404.3 ~\mu K^2$).
We conclude that the $C_2$ probability of increasing (decreasing) is $69 \%$ ($31 \%$).

\end{document}